\documentclass[12pt]{article}

\usepackage{amsmath}

\begin{document}

\title{Scientific Methodology: A View from Early String Theory\thanks{Forthcoming in:
 R. Dawid, R. Dardashti, and K. Th\'ebault. {\it Epistemology of fundamental physics: Why trust a theory?}, Cambridge: Cambridge University Press.}}

\author{Elena Castellani\thanks{Department of Humanities and Philosophy, University of Florence, via Bolognese 52, 50139,
Firenze, Italy. E-mail: elena.castellani@unifi.it.}}

\date{January 2018}

\maketitle

\section{Introduction}

This contribution is devoted to addressing the question as to whether the methodology followed in building/assessing string theory can be considered ÔscientificÕ -- in the same sense, say, that the methodology followed in building/assessing the Standard Model of particle physics is scientific -- by focussing on the "founding" period of the theory. More precisely, its aim is to argue for a positive answer to the above question in the light of a historical analysis of the early developments of the string theoretical framework.

The paper's main claim is a simple one: there is no real change of scientific status in the way of proceeding and reasoning in fundamental physical research. Looking at the developments of quantum field theory and string theory since their very beginning, one sees the very same strategies at work both in theory building and theory assessment. Indeed, as the history of string theory clearly shows (see Cappelli et al., 2012), the methodology characterising the theoretical process leading to the string idea and its successive developments is not significantly different from the one characterising many fundamental developments in theoretical physics which have been crowned with successful empirical confirmation afterwards (sometimes after a considerable number of years, as exemplified by the story of the Higgs particle). 

Of course, theory assessment based on empirical support is problematic in the case of theoretical developments that concern regimes (energy/length scales) away from the current possibility of empirical access, such as those related to the string theoretical framework. This happens for reasons that can be both technical and theoretical -- the technology is not advanced enough or the theory is not developed enough. In such cases, non-empirical criteria naturally play a major role in theory assessment (which does not mean renouncing to pursue empirical confirmation, it is worth stressing).

The stance adopted here is that the real issue at stake, when discussing the status of string theory, is not whether non-empirical theory assessment could or should substitute the traditional way of confirming physical theories by confrontation with empirical data. To understand the debate in such terms, I think, is misleading and fruitless.\footnote{On the fact that a physical theory must receive, sooner or later, an empirical confirmation -- in whatever way - there can be no real disagreement.}
The genuine question is rather how to obtain a reasonable balance between empirical and non-empirical criteria for theory assessment, given the particular physical context considered. From the methodological point of view, the only real difference is how this balance is achieved and justified, depending on the specific scenarios one is dealing with.  Thus, in the case considered here, the main issue is: what are the admissible typology and extent of non-empirical support that can reasonably motivate pursuing the chosen line of research? 

Criteria and strategies for this kind of support to theory assessment are of various types. While the ``non-empirical confirmation arguments'' individuated by Dawid (2013) -- the Ê``No AlternativesÓ Argument'', the ``Meta-inductive Argument'' and the ``Unexpected-Explanation Argument''  -- work at a general epistemological `meta-level', for what concerns the `object-level' of scientific practice the philosophical discussion has been traditionally focused on the so-called theoretical virtues: that is, virtues such as simplicity, coherence and elegance, to mention some of the main ones.

Here, we will focus on a further type of criterium influential at the level of scientific practice. This is the criterium grounded on what can be called the ``Convergence Argument'': namely, the argument for non-empirical support which is based on the fact that some of the crucial new results are obtained in alternative, independent ways, and even from different starting points.\footnote{Let us note that this criterium could be seen to have some affinity with the famous ``consilience of inductions" introduced, as one of the confirmation criteria (beside predictivity and coherence), by William Whewell in his 1840  {\it The Philosophy of Inductive Sciences, Founded Upon Their History}. This point is examined in another paper in preparation, in collaboration with Radin Dardashti and Richard Dawid.}
This sort of argument has been particularly effective in providing motivation for accepting apparently very unphysical features emerging from theoretical developments. A paradigmatic example is given by the story of the discovery of extra dimensions (22 extra space dimensions) in the framework of ``early string theory'', an expression used to indicate the origin and first developments of string theory, from Veneziano's 1968 formulation of his famous scattering amplitude to the so-called first string revolution in 1984.\footnote{See Cappelli et al., 2012, for detailed historical and, in many cases, first-hand  reconstructions of these early developments of string theory.}

In what follows, after some preliminary remarks about the nature and merit of the methodological debate concerning what is currently known as string theory, we will focus on early string theory and its significance for the issue at stake. More precisely, the focus will be on the first phase of this `founding era' of the theory. 
The aim is to show, by looking in some detail at the extra dimension case, that the rationale guiding both the building and assessment methodology in cases like this one -- where features far from experimental testing possibility are introduced and become a basic ingredient of the whole theoretical framework -- is not significantly different from the methodology followed in many successful theoretical advances in physics, later empirically confirmed.

\section{Scientific methodology and string theory: the issue} 

Traditionally, the general methodological debate in the philosophy of science regards:

\begin{itemize}

\item[$\bullet$] the modalities and strategies followed in building scientific theories, i. e.  questions about scientific heuristic or `discovery';

\item[$\bullet$] the modalities and strategies followed in assessing/confirming
scientific theories (on the grounds of both empirical and extra-empirical support), i. e. questions about `justification'; 

\item[$\bullet$] the inter-relations between theory building and theory assessing, and the influence of `external' aspects (psychological, sociological, economical, political) in the scientific enterprise.

\end{itemize}

The 2015 Munich Workshop on ``Why Trust a Theory?'', on which this volume is based,  was substantially concerned with the second point (with some attention to  the influence of sociological aspects on research strategies, in addition).  
In fact, the specific question was whether the methodology followed in assessing string theory (and other theories that are currently in a similar situation as regard empirical support) could be said {\it scientific} -- which is not the same thing, of course, as asking whether string theory is {\it true}.

Indeed, methodological queries regarding string theory are generally concerned with justification, not discovery. 
There is a sort of implicit consensus that the problem doesn't lie specifically in the theory building phase.
What is considered controversial is the scientific status of the results of such a building activity, because of the lack of empirical confirmation. 

However, assuming a clear-cut distinction between the two modalities of
scientific activity -- model/theory building and assessment strategies -- is not
unproblematic, as is well known in philosophy of science.\footnote{A detailed revisitation of this distinction (its merits and its flaws), also in the light of its history, is provided in the collective volume edited by Schickore and Steinle, 2006.} In the current debate on the status of string theory, 
moreover, the actual disagreement is 
on the use of non-empirical criteria in theory assessment, not on the
modalities to be followed for an empirical confirmation. The real issue, in fact,  
is the nature and legitimacy of the non-empirical criteria, to be considered in the light of the role they play
in the intertwining of discovery/assessing strategies.

As already mentioned in Section 1,  two sorts of criteria for extra-empirical support to scientific theories can be distinguished:

\begin{itemize}

\item[$\bullet$] General meta-criteria, implying the confrontation of the theory considered with other theories. Best examples of such  `external' criteria are the three ``non-empirical confirmation arguments'' individuated by Dawid (2013), much debated in this volume.

\item[$\bullet$]  More specific `internal' criteria, based on those theoretical virtues or `values' which are typically discussed in philosophy of science with respect to the issues of theory underdetermination and theory choice. An inventory of such values traditionally includes, beside the already mentioned simplicity, coherence and elegance (or beauty), also consistency, unifying power/generality, fertility and explanatory power.\footnote{See for example Ivanova and Paternotte, 2017, for an overview of the discussion of such kinds of virtues in evaluating theories. Historically, this discussion was particularly boosted by Kuhn's view on the role of values in theory choice, defended in his reply to his critics in Lakatos and Musgrave, 1970. Kuhn's answer to those who criticised him for making a case for relativism is notoriously based on a list of five values that, according to him, were used by scientists in the choice of one theory over another: namely, accuracy, consistency, scope, simplicity, and fertility.}

\end{itemize}

As said, other types of internal criteria can be individuated. Here, we will consider the role of one of them, the {\it convergence argument}, by discussing a specific case study: namely, the discovery of extra dimensions in the context of early string theory.

\section{A case study from early string theory}

 Early string theory (EST) represents a particularly fruitful case for discussing the modalities of theory building and theory assessment in fundamental physics. Indeed, EST's history provides light on the origin of many ideas (string, duality, supersymmetry, extra dimensions, ...) and mathematical techniques that have become basic ingredients in this research field. Moreover, by highlighting the rationale of a scientific progress characterised by a close interplay of mathematically driven creativity and physical constraints, the historical analysis of these developments do provide novel data for discussing the building/assessing issue and, in particular, the role and characteristics of the evidential support  in the construction process of a scientific theory.

Here, we will focus on the first phase of this founding era of string theory: the so-called {\it dual theory of strong interactions}, flourishing in the years 
1968-1973.\footnote{On the distinction between two phases of EST - a first phase lasting till the end of 1973, when EST was falsified as a theory of strong interactions, and a second phase of circa a decade (1974-1984), where the theory was re-interpreted as  a unified quantum theory of all fundamental interactions  -- see Castellani,  in Cappelli et al., 2012, Chapter 4, section 1. More detailed descriptions are contained in Cappelli et al., 2012, Part 1.} This is the period in which, following Veneziano's 1968 discovery of his ``dual amplitude'' for the scattering of four mesons, a very intense model building activity developed, from the first two models for the scattering of $N$ particles -- the generalised Veneziano model, known as the {\it Dual Resonance Model} (DRM), and the {\it Shapiro-Virasoro Model}\footnote{These two models were later understood as describing open and closed strings, respectively.} -- to all the subsequent endeavours to extend, complete and refine the theoretical framework, including its string interpretation and the addition of fermions.

As is well known, this first phase of early string theory was originally aimed at finding a viable theory of hadrons in the framework of the so-called analytic $S$-matrix (or $S$-matrix theory) developed in the early Sixties.\footnote{On the $S$-matrix programme pursued by Chew and his collaborators, see in particular Cushing, 1990, and Cappelli et., 2012, Part II.} The $S$-matrix programme, initially pursued in particular by Geoffrey Chew and his collaborators, was motivated by the difficulties arising in a field-theoretic description of strong interactions. Inspired by earlier works of Wheeler and Heisenberg, its aim was to determine the relevant observable physical quantities, i.e. the scattering amplitudes, only on the basis of some general principles such as {\it unitarity}, {\it analiticity} and {\it crossing symmetry} and a minimal number of additional assumptions. 

One of these assumptions, suggested by experimental data, was the so-called {\it duality principle}, introduced in 1967 by Dolen, Horn and Schmid. The meaning of this duality, also known as {\it DHS duality}, was that 
the contributions from resonance intermediate states and from particle exchange each formed a complete representation of the
scattering process (so that they should not be added to one another in
order to obtain the total amplitude).\footnote{More precisely, in terms of Mandelstam's variables and using the framework of the so-called Regge theory,  the duality principle (as initially stated) established direct relations between a low-energy and a high-energy description of the hadronic scattering amplitude $A(s,t)$: namely, the low-energy description  in terms of direct-channel ($s$-channel) resonance poles, and the high-energy description in terms of the exchange of Regge poles in the crossed-channel ($t$-channel), could each be obtained from the other by analytic continuation. See Cappelli et al., 2012, Part II, 5.4.3, for details.} In fact, the duality principle was seen as representing an effective implementation of two connected ideas defended, in particular,
by Chew and his school: on the one hand, the idea of ``nuclear democracy'', according to which no hadron is
more fundamental than the others; on the other hand, the ``bootstrap idea'', that is, the idea of a self-consistent hadronic
structure in which the entire ensemble of hadrons provided the forces, by hadron exchange, making their own existence possible.\footnote{Whence, also, the name of {\it dual bootstrap} for the DHS duality.}

In 1968, Gabriele Veneziano achieved to obtain a first, brilliant solution to the problem of finding, in the
framework of the $S$-matrix approach, a scattering amplitude obeying also the duality
principle. This ground-breaking result is officially recognised as the starting point of early string theory. A period of intense theoretical activity immediately followed Veneziano's discovery, aimed at extending his dual amplitude in order to overcome its limits: first of all, the fact that only four particles, and of a specific type (mesons), were considered. Moreover, the model violated unitarity because of the narrow-resonance approximation.\footnote{That is, the approximation corresponding to the fact that, in each channel, only single-particle exchanges were considered; see on this Ademollo, in Cappelli et al., 2012, Chapter 6, Section 6.7.} It was also natural to search for different models that could include other neglected but important physical features (such as fermionic particles, beside bosons), and, more generally, to try to reach a better understanding of the physical theory underlying the models that were being constructed.

This theory building process was characterised by two particularly significant conjectures. 
First of all, the string conjecture in 1969: in independent attempts to gain a deeper understanding of the physics described by dual amplitudes, 
Nambu, Nielsen and Susskind each arrived at the conjecture that the underlying dynamics of the dual resonance model
was that of a quantum-relativistic oscillating string. Second, the conjecture or `discovery' of extra spacetime dimensions: consistency conditions 
in the developments of the dual theory led to the critical value $d=26$ for the spacetime dimension (known as the {\it critical dimension}), reducing to the value $d=10$ when including fermions.  

Also in this second case, the conjecture was arrived at by following independent paths, although under the same general motivation -- that is, to extend the original dual theory and overcome its initial limitations and problems.  
The result was a bold conjecture -- 22 extra space dimensions -- and there is no doubt that the fact that it was obtained in different, independent ways (and even from  different starting points) was an influential reason for taking it seriously. In what follows, we will briefly outline this significant discovery case, illustrative of both the
rationale leading to apparently bold guesses  and the kind of evidential support motivating a theory's progress.

\subsection{Ways to the ``critical dimension''}

In the first 
years of early string theory, we can distinguish three different approaches to arrive at the critical dimension.
The first two ways have more to do with the building/discovery process, the third one also with theory assessment. 

Originally, the critical dimension emerged in the context of two independent programmes in the framework of the Dual Resonance Model (DRM): namely, (1) the ``unitarisation programme''  and (2) the ``ghost elimination programme''.

\begin{itemize} 

\item[(1)] In the first case, Claud Lovelace 
arrived at the conjecture $d=26$ while addressing a problematic singularity case arising in the construction of the nonplanar one-loop amplitude.

\item[(2)] In the second case, the critical value $d=26$ for the spacetime dimension 
 issued from studying the spectrum of states of the Dual Resonance Model. 
 
 \end{itemize}
 
Let us have a closer look at these first two ways, keeping the details to a minimum.

\subsubsection{Lovelace's way (1971)}

As mentioned above, the original dual amplitudes didn't respect the $S$-matrix unitarity condition.
To go beyond the initial narrow-resonance approximation, the ``unitarization programme''  was based on the analogy between this approximation in the dual theory 
and the so-called Born approximation (or ``tree approximation'') in conventional quantum field theory. The programme was thus to generalise the
initial amplitudes, considered as the lowest order or tree diagrams of a perturbative expansion, 
to include loops. As a first step for restoring unitarity, one-loop diagrams were constructed, and in this building process the calculation of
a nonplanar loop diagram led to the conjecture of the value $d=26$ for the spacetime dimension. 
This result was obtained by Claud Lovelace in a  work published in 1971.

In some more detail: Lovelace arrived at the so-called {\it critical dimension} $d=26$ by addressing a singularity problem emerged in the construction 
of the nonplanar one-loop amplitude.\footnote{In four spacetime dimensions, the amplitude had a singularity (a `branch cut') in a certain channel, incompatible with unitarity. See Lovelace's own account of his `discovery' in Cappelli et al., 2012, Chapter 15.} He realised that the singularity could be turned into a {\it pole}, and thus interpreted as due to the propagation of a 
new intermediate particle, if the value of the spacetime dimension was $d=26$. 
This pole, Lovelace conjectured to be the Pomeron, 
the particle that was later understood as the graviton.\footnote{For details, see Cappelli et al., 2012, Section 10.2.3.} 
The decisive step, indeed, was to consider the possibility 
that the spacetime dimension $d$ might be different from $4$ and treat it as
a free parameter.\footnote{This was not the only condition, but there is no need to enter into such details, here. 
But let's remember, for historical sake,  Olive 's personal account of Lovelace's discovery (Cappelli et al., Chapter 28, 398): ``In late June 1970, I
remember Claud Lovelace giving a talk about his evaluation of the twisted
loop amplitude in four dimensions. He had found what sounded like a disaster
in that the result of a precise calculation revealed a branch point singularity
in a certain channel. I explained to him that if only the singularity
could be a simple pole, rather than a branch cut, it could be consistent with
general principles of scattering matrix theory, such as unitarity, as it could
be interpreted as being due to the propagation of a new sort of particle (I
have to say that Claud has denied all memory of this episode when I tried
to check it with him in December 2007)''.}

 A spacetime of 26 dimensions was not easy to accept, especially in the context of the phenomenology of strong interactions 
 where it was proposed. 
 In a recollection paper on his contribution to the dual theory, Lovelace describes 
the first reactions to his conjecture as follows:  ``I gave a seminar ... which was attended by
some powerful people as well as the Dual Model group. Treating the result
as a joke, I said I had bootstrapped the dimension of spacetime but the
result was slightly too big. Everyone laughed.''\footnote{Cappelli et al, 2012, Chapter 15, 228.}. As he himself acknowledges, 
one had to be ``very brave to suggest that spacetime has 26 dimensions''.
 
However, almost at same time, Lovelace's ``wild conjecture'' (his words) was vindicated through another completely independent route:
 the very same number of spacetime dimensions made its appearance in the context of the ghost elimination programme.

\subsubsection{The ``no ghost'' way}

Soon after Veneziano's discovery of his amplitude for the scattering of four
scalar particles, endeavours started for its generalisation to the scattering of an arbitrary number $N$ of scalar particles. 
In the following studies of the properties of the resulting model, the multi-particle Veneziano model known as the Dual Resonance
Model, a serious problem was represented by the presence of 
negative-norm states -- or ``ghosts'', as they were called at the time --\footnote{Note that this is a different meaning of the term ``ghost'' with respect to how it is commonly used in quantum field theory (i.e., to indicate the unphysical fields associated with gauge invariance in functional approaches to field theory quantisation).}  in the state spectrum of the model. Such kinds of states, leading 
to unphysical negative probabilities, had to be eliminated from the theory. 

The strategy adopted for the ghost elimination programme was based on an analogy suggested by
a similar situation encountered in the covariant quantisation of electrodynamics, where the unphysical negative-norm states 
were removed by using the gauge invariance of the theory 
and the  ``Fermi condition'' following from it  (the equation characterizing the positive-norm physical states).
The DRM analogues of the conditions imposed by the gauge invariance were found to be given by the so-called Virasoro conditions --
an infinite number of relations later understood as associated with the infinite-dimensional symmetry corresponding to
the {\it conformal transformations} of the two-dimensional string world-sheet.\footnote{See, in particular, Di Vecchia, in Cappelli et al., 2012, Sections 11.3 and 11.4.} 

At that point, a further step towards the elimination of the unwanted ghost states 
was the 1971 construction by Del Giudice, Di Vecchia and Fubini of an infinite set of positive-norm states, known as the {\it DDF states}. 
Initially, though, these states did not seem to be sufficient to span the whole Hilbert space.
But already one year after, the result was obtained that the DDF states could indeed span the whole space of physical states if the spacetime dimension $d$ was equal to 26  -- the very same value as the one conjectured by Lovelace. 
Soon after, the proof of the so-called No-Ghost Theorem, establishing that the Dual
Resonance Model has no ghosts if $d \leq 26$, was achieved independently by Brower  and by Goddard and Thorn.\footnote{By essentially same argument as in the case of the DRM, it was also proved that Neveu-Schwarz dual model has no ghosts if $d \leq 10$, thus confirming the critical dimension as $d = 10$ in the case including fermions. A detailed description of the No-Ghost result can be found, in particular, in Goddard's contribution to Cappelli et al., 2012, Chapter 20.}
 
While initially almost nobody had taken Lovelace's conjecture seriously, after the proof
of the No-Ghost Theorem the attitude changed and the extra dimensions started to be gradually accepted in the dual model community.\footnote{A good example is given in the following quote by Goddard (Cappelli et al., 2012, Chapter 20,  285): ``The validity of the No-Ghost Theorem had a profound effect on me. It seemed clear that this result was quite a deep mathematical statement 
..., but also that no pure mathematician would have written it down.
It had been conjectured by theoretical physicists because it was a necessary
condition for a mathematical model of particle physics not to be inconsistent
with physical principles. ... I could not help thinking that, in some sense, there would be
no reason for this striking result to exist unless the dual model had something
to do with physics, though not necessarily in the physical context in which
it had been born.''}
A further decisive support came from an immediately successive theoretical
development, leading, among other things, to the same `critical' value $d = 26$ for the spacetime dimension: namely, 
the 1973 work of Goddard, Goldstone, Rebbi and Thorn (GGRT) on the quantisation of
the string action.  

\subsubsection{The GGRT (1973) way}

As we have seen, the $S$-matrix approach was based on the construction of observable scattering amplitudes. 
Nonetheless, soon after Veneziano's result and its first generalisations, a physical interpretation of the dual amplitudes in terms of an underlying dynamics and an appropriate Lagrangian started to be investigated. Already in 1969, Nambu, Nielsen and Susskind each independently made the conjecture that the dynamics of the dual resonance model was that of an oscillating string. And in the following year Nambu (and then Goto) proposed the Lagrangian action for the string, 
 formulated in terms of the area of the surface swept out by a one-dimensional extended object moving in spacetime, in analogy with the formulation of the action of a point particle in terms of the length of its trajectory.  
 
It took some time, however, for the string interpretation of the dual resonance model to be fully accepted. It was not clear, originally, whether 
 the conjecture was a mere analogy,\footnote{The analogy between the structure of the DRM spectrum and that of a vibrating string was based on the harmonic oscillator: on the one hand, the DRM states were described in terms of an infinite number of creation and annihilation operators of the harmonic oscillator; on the other hand, a vibrating string could be described as an infinite collection of harmonic oscillators where the harmonics are multiples of a fundamental frequency.}  useful for calculations, or it had deeper physical meaning. The work by Goddard, Goldstone, Rebbi and Thorn on the quantisation of the string action had a decisive impact, in this respect: thanks to their result, the string hypothesis became an essential aspect of the theory, revealing its underlying structure. The analogy thus turned out to provide an effective `interpretation', playing an influential role in the transformation process from the original dual models to the string theoretical framework.

In the resulting description, all what had been previously obtained by proceeding according to a bottom-up approach and via various routes could now be derived in a more clear and unitary way. In particular, for what regards the critical dimension, it was obtained as a condition for the Lorentz invariance of the canonical quantisation of the string in the light-cone gauge: only for $d=26$ the quantisation procedure was Lorentz invariant.\footnote{Details on this point, and in general on the quantisation of the hadronic string, are described by Di Vecchia and Goddard in their contributions to Cappelli et al., 2012, Chapter 11 (11.8) and Chapter 20 (20.7), respectively.}
Thus, in a certain sense, the GGRT way to arrive at the extra dimensions had both to do with discovery and theory assessment.

\section{Conclusion}

With respect to the acceptance of the extra dimension conjecture in the string theory community, the three independent ways to arrive at the one and  same result illustrated above are not the whole story, of course.  
Further and more decisive support to this conjecture came from successive developments of string theory, especially 
after it was re-interpreted as a unified quantum theory of all fundamental interactions including gravity.\footnote{See Cappelli et al., 2012, Part VI.} In fact, today's way of understanding the critical dimension is in terms of what is technically called an `anomaly': 
as shown in seminal 1981 work by Polyakov in the framework of his path-integral approach to the quantisation of the string action,
the conformal symmetry of the classical string Lagrangian is `anomalous', i.e. not conserved 
in the quantum theory, unless the value of the spacetime dimension is $d = 26$ ($d = 10$, in the case with fermions).\footnote{Polyakov, in his contribution to Cappelli et al., 2012, Chapter 44, gives his own personal recollection of these developments; for a more general overview, see also Cappelli et al, Part VII, Chapter 42.} 

Thus, with hindsight, the convergence of the different, independent ways to obtain the critical dimension
can be understood as one of the remarkable consequences of the strong constraints put on the theory by its conformal symmetry. 
At this point, one could object that to apply here the `convergence argument' as a criterium for theory assessment appears a bit circular. It looks like 
deriving (extra-empirical) support to the extra dimension conjecture on the grounds of another essential feature of the theory itself. 

Let us say two things in this regard. First, what we are interested in, in this article, is the role of extra empirical criteria -- like the convergence argument -- in guiding the building of a theory and motivating its acceptance. The case of extra dimensions in early string theory surely illustrates this role. 
At the same time, to enter into the details of this case study allows us to show, from a methodological point of view, the concrete procedures followed in the  development of the theoretical framework: originally a bottom-up activity, starting from a very phenomenological basis and successively progressing {\it via} generalisations, conjectures and analogies, suggested and constrained by physical principles and data. A very typical example of scientific methodology.

Second and finally: even if, on the one side, the convergence to the same surprising result of independent ways of investigation can be explained, 
afterwards, as a natural consequence of the more mature theoretical framework -- thus implying some circularity when using this fact as an assessment criterium for a full-fledged theory -- this very fact, on the other side, speaks in favour of `virtues' of the description obtained
such as its cohesiveness and consistency. In this case, we can say, the convergence criterium gives place to the other form of `internal' criteria mentioned in Section 2, that is the one based on the theoretical virtues influential in theory assessment.

\bigskip

\begin{center} 
{\bf Acknowledgements}
\end{center}

\noindent I am very grateful to the organisers and the audience of the "Why
Trust a Theory?" workshop in Munich. A special thanks to Erik Curiel, Radin Dardashti and Richard Dawid for insightful and helpful discussions, and to the Munich Centre for Mathematical Philosophy for a visiting fellowship in December 2015, 
allowing me to enjoy its great and collaborative atmosphere.

\begin{center}
{\bf References}
\end{center}

\noindent R. C. Brower (1972), ``Spectrum-generating algebra and no-ghost theorem in
the dual model'', {\it Phys. Rev.} D6: 1655-1662.\\

\noindent A. Cappelli, E. Castellani, F. Colomo, and P. Di Vecchia (Eds.) (2012), {\it The Birth of String Theory}, Cambrige: Cambridge University Press.\\

\noindent J. T. Cushing (1990), {\it Theory Construction and Selection in Modern Physics: The S-Matrix}, Cambridge: Cambridge University Press.\\

\noindent R. Dawid (2013), {\it String Theory and the Scientific Method}, Cambridge:
Cambridge University Press.\\

\noindent P. Goddard and C.B.Thorn (1972), ``Compatibility of the pomeron with
unitarity and the absence of ghosts in the dual resonance model'', {\it Phys. Lett.}
B40: 235-238.\\

\noindent P. Goddard, J. Goldstone, C. Rebbi, and C. B. Thorn (1973), ``Quantum
dynamics of a massless relativistic string'', {\it Nucl. Phys.} B56: 109-135.\\

\noindent M. Ivanova and C.Paternotte (2017), ``Virtues and vices in scientific practice'',  {Synthese} 194: 1787-1807.\\

\noindent T. Kuhn, ``Reflections on my Critics'' (1970), in I. Lakatos and A.Musgrave (Eds.), {\it Criticism and the Growth of Knowledge}, Cambridge: Cambridge University Press, pp. 231-278.\\

\noindent C. Lovelace (1971), ``Pomeron form factors and dual Regge cuts'', {\it Phys. Lett.} B34: 500-506.\\

\noindent J. Schickore and F. Steinle (Eds.) (2006), {\it Revisiting Discovery and Justification. Historical and philosophical perspectives on the context distinction}, Dordrecht: Springer.\\

\noindent G. Veneziano (1968),  ``Construction of a crossing-symmetric, Reggeon behaved
amplitude for linearly rising trajectories'', {\it Nuovo Cimento} A57: 190-197.

\end{document}